\documentclass{elsarticle}

\usepackage{amsmath}
\usepackage[dvipsnames]{xcolor}
\usepackage{subfig}
\usepackage{multirow}
\usepackage{hyperref}
\usepackage{algorithm}
\usepackage{graphicx}
\usepackage{diagbox}
\usepackage{caption}
\usepackage{subfig}
\usepackage{longtable}
\usepackage{threeparttable}
\usepackage{comment}
\usepackage{textcomp}

\journal{Medical Image Analysis}

\newcommand{\red}[1]{\textcolor{red}{#1}}
\newcommand{\redb}[1]{\textcolor{red}{[#1]}}
\newcommand{\blue}[1]{\textcolor{blue}{#1}}

\graphicspath{{./figs/}}

\begin{document}

\begin{frontmatter}

\title{Integrative Analysis for COVID-19 Patient Outcome Prediction}

\author[rpi]{Hanqing Chao\corref{cofirstauthor}}
\author[rpi]{Xi Fang\corref{cofirstauthor}}
\author[rpi]{Jiajin Zhang\corref{cofirstauthor}}

\author[mgh]{Fatemeh~Homayounieh}
\author[mgh]{Chiara~D.~Arru}
\author[mgh]{Subba R. Digumarthy}

\author[Firoozgar]{Rosa Babaei}
\author[Firoozgar]{Hadi K. Mobin}
\author[Firoozgar]{Iman Mohseni}

\author[Cagliari]{Luca Saba}
\author[Novara]{Alessandro Carriero}
\author[Novara]{Zeno Falaschi}
\author[Novara]{Alessio~Pasche}

\author[rpi]{Ge~Wang}

\author[mgh]{Mannudeep~K.~Kalra\corref{mycorrespondingauthor}}

\author[rpi]{Pingkun~Yan\corref{mycorrespondingauthor}}

\cortext[cofirstauthor]{Equally contributed first authors}
\cortext[mycorrespondingauthor]{Co-corresponding authors}
\ead{MKALRA@mgh.harvard.edu, yanp2@rpi.edu}

\address[rpi]{Department of Biomedical Engineering and the Center for Biotechnology and Interdisciplinary Studies at Rensselaer Polytechnic Institute, Troy, NY 12180, USA.}
\address[mgh]{Department of Radiology, Massachusetts General Hospital, Harvard Medical School, Boston MA 02114, USA}
\address[Firoozgar]{Department of Radiology, Firoozgar Hospital, Iran University of Medical Sciences, Tehran, Iran}
\address[Cagliari]{Azienda Ospedaliero-universitaria di Cagliari, Cagliari, Italy}
\address[Novara]{Azienda Ospedaliera Ospedale Maggiore della Carita' di Novara, Novara, Italy}
\begin{abstract}
While image analysis of chest computed tomography (CT) for COVID-19 diagnosis has been intensively studied, little work has been performed for image-based patient outcome prediction. Management of high-risk patients with early intervention is a key to lower the fatality rate of COVID-19 pneumonia, as a majority of patients recover naturally. Therefore, an accurate prediction of disease progression with baseline imaging at the time of the initial presentation can help in patient management.
In lieu of only size and volume information of pulmonary abnormalities and features through deep learning based image segmentation, here we combine radiomics of lung opacities and non-imaging features from demographic data, vital signs, and laboratory findings to predict need for intensive care unit (ICU) admission. To our knowledge, this is the first study that uses holistic information of a patient including both imaging and non-imaging data for outcome prediction.
The proposed methods were thoroughly evaluated on datasets separately collected from three hospitals, one in the United States, one in Iran, and another in Italy, with a total 295 patients with reverse transcription polymerase chain reaction (RT-PCR) assay positive COVID-19 pneumonia. Our experimental results demonstrate that adding non-imaging features can significantly improve the performance of prediction to achieve AUC up to 0.884 and sensitivity as high as 96.1\%, which can be valuable to provide clinical decision support in managing COVID-19 patients. Our methods may also be applied to other lung diseases including but not limited to community acquired pneumonia. The source code of our work is available at \url{https://github.com/DIAL-RPI/COVID19-ICUPrediction}.
\end{abstract}


\end{frontmatter}

\section{Introduction} 
\label{intro}

Coronavirus disease 2019 (COVID-19), which results from contracting an extremely contagious beta-coronavirus, is responsible for the latest pandemic in human history. The resultant lung injury from COVID-19 pneumonia can progress rapidly to diffuse alveolar damage, acute lung failure, and even death \citep{vaduganathan_2020,danser_2020}. Given the highly contagious nature of the infection, the burden of COVID-19 pneumonia has imposed substantial constraints on the global healthcare systems.
In this paper, we present a novel framework of integrative analysis of heterogeneous data including not only medical images, but also patient demographic information, vital signs and laboratory blood test results for assessing disease severity and predicting intensive care unit (ICU) admission of COVID-19 patients. Screening out the high-risk patients, who may need intensive care later, and monitoring them more closely to provide early intervention may help save their lives. 

Reverse transcription polymerase chain reaction (RT-PCR) assay with detection of specific nuclei acid of SARS-CoV-2 in oral or nasopharyngeal swabs is the preferred test for diagnosis of COVID-19 infection. Although chest computed tomography (CT) can be negative in early disease, it can achieve higher than 90\% sensitivity in detecting COVID-19 pneumonia but with low specificity \citep{kim_diagnostic_2020}. For diagnosis of COVID-19 pneumonia, CT is commonly used in regions with high prevalance and limited RT-PCR availability as well as in patients with suspected false negative RT-PCR. CT provides invaluable information in patients with moderate to severe disease to assess the severity and complications of COVID-19 pneumonia \citep{yang_chest_2020}. Prior clinical studies with chest CT have reported that qualitative scoring of lung lobar involvement by pulmonary opacities (high lobar involvement scores) can help assess severe and critical COVID-19 pneumonia. \cite{li_clinical_2020} showed that high CT severity scores (suggestive of extensive lobar involvement) and consolidation are associated with severe COVID-19 pneumonia. \cite{zhao_relation_2020} reported that extent and type of pulmonary opacities can help establish severity of COVID-19 pneumonia. The lung attenuation values change with the extent and type of pulmonary opacities, which differ in patients with more extensive, severe disease from those with milder disease.
Most clinical studies focus on qualitative assessment and grading of pulmonary involvement in each lung lobe to establish disease severity, which is both time-consuming and associated with interobserver variations \citep{zhao_relation_2020,ai_correlation_2020}. 
To address the urgent clinical needs, artificial intelligence (AI), especially deep learning, has been applied to COVID-19 CT image analysis \citep{shi_review_2020}. AI has been used to differentiate COVID-19 from community acquired pneumonia (CAP) on chest CT images \citep{li_artificial_2020,sun_adaptive_2020}. To unveil what deep learning uses to diagnose COVID-19 from CT, \cite{wu_jcs_2020} proposed an explainable diagnosis system by classifying and segmenting infections. \cite{gozes_coronavirus_2020} developed a deep learning based pipeline to segment lung, classify 2D slices and localize COVID-19 manifestation from chest CT scans. \cite{shan_lung_2020} went on to quantify lung infection of COVID-19 pneumonia from CT images using deep learning based image segmentation.  

Among the emerging works, a few AI based methods target at severity assessment from chest CT. \cite{huang_serial_2020} developed a deep learning method to quantify severity from serial chest CT scans to monitor the disease progression of COVID-19. \cite{tang_severity_2020} used random forest to classify pulmonary opacity volume based features into four severity groups. By automatically segmenting the lung lobes and infection areas, \cite{gozes_rapid_2020} suggested a ``Corona Score'' to measure the progression of disease over time. \cite{zhu_joint_2020} further proposed to use AI to predict if a patient may develop severe symptoms of COVID-19 and how long it may take if that is the case. Although promising results have been presented, the existing methods primarily focus on the volume of pulmonary opacities and their relative ratio to the lung volume for severity assessment. The type of pulmonary opacities (e.g. ground glass, consolidation, crazy-paving pattern, organizing pneumonia) is also an important indicator of the stage of the disease and is often not quantified by the AI algorithms \citep{chung_ct_2020}.

Furthermore, in addition to measuring and monitoring the progression of severity, it could be life-saving to predict mortality risk of patients by learning from the clinical outcomes. Since majority of the infected patients will recover, managing the high-risk patients is the key to lower the fatality rate \citep{Ruan2020, Phua2020, Li2020}. Longitudinal study analyzing the serial CT findings over time in patients with COVID-19 pneumonia shows that the temporal changes of the diverse CT manifestations follow a specific pattern correlating with the progression and recovery of the illness \citep{wang_temporal_2020}. Thus, it is promising for AI to perform this challenging task.

In this paper, our objective is to predict outcome of COVID-19 pneumonia patients in terms of the need for ICU admission with both imaging and non-imaging information. The work has two major contributions.
\begin{enumerate}
	\item  While image features have been commonly exploited by the medical image analysis community for COVID-19 diagnosis and severity assessment, non-imaging features are much less studied. However, non-imaging health data may also be strongly associated with patient severity. For example, \cite{yan_interpretable_2020} showed that machine learning tools using three biomarkers, including lactic dehydrogenase (LDH), lymphocyte and high-sensitivity C-reactive protein (hs-CRP), can predict the mortality of individual patients. Thus, we propose to integrate heterogeneous data from different sources, including imaging data, age, sex, vital signs, and blood test results to predict patient outcome. To the best of our knowledge, this is the first study that uses holistic information of a patient including both imaging and non-imaging data for outcome prediction. 
	\item In addition to the simple volume measurement based image features, radiomics features are computed to describe the texture and shape of pulmonary opacities. A deep learning based pyramid-input pyramid-output image segmentation algorithm is used to quantify the extent and volume of lung manifestations. A feature dimension reduction algorithm is further proposed to select the most important features, which is then followed by a classifier for prediction.
	
\end{enumerate}
It is worth noting that although the presented application on COVID-19 pneumonia, the proposed method is a general approach and can be applied to other diseases.

The proposed method was evaluated on datasets collected from teaching hospitals across three  countries, These datasets included 113 CT images from Firoozgar Hospital (Tehran, Iran)(Site A), 125 CT images from Massachusetts General Hospital (Boston, MA, USA)(Site B), and 57 CT images from University Hospital Maggiore della Carita (Novara, Piedmont, Italy)(Site C). Promising experimental results for outcome prediction were obtained on all the datasets with our proposed method, with reasonable generalization across the datasets. Details of our work are presented in the following sections.

\section{Datasets}
\label{sec:ds}
The data used in our work were acquired from three sites. All the CT imaging data were from patients who underwent clinically indicated, standard-of-care, non-contrast chest CT without intravenous contrast injection. Age and gender of all patients were recorded. For datasets from Sites A and B, lymphocyte count and white blood cell count were also available. For datasets of Sites A and C, peripheral capillary oxygen saturation (SpO2) and temperature on hospital admission were recorded. Information pertaining patient status (discharged, deceased, or under treatment at the time of data analysis) was also recorded as well as the number of days of hospitalization to the outcome.  

\paragraph{Site A Dataset} We reviewed medical records of adult patients admitted with known or suspected COVID-19 pneumonia in Firoozgar Hospital (Tehran, Iran) between February 23, 2020 and March 30, 2020. Among the 117 patients with positive RT-PCR assay for COVID-19, three patients were excluded due to presence of extensive motion artifacts on their chest CT. With one patient who neither admitted to ICU nor discharged, 113 patients are used in this study.

\paragraph{Site B Dataset} We reviewed medical records of adult patients admitted with COVID-19 symptom in MGH between March 11 and May 3, 2020. 125 RT-PCR positive admitted patients underwent unenhanced chest CT are selected to form this dataset.

\paragraph{Site C Dataset} We reviewed medical records of adult patients admitted with COVID-19 pneumonia in the Novara Hospital (Piedmont, Italy) between March 4, 2020 and April 6, 2020. We collected clinical and outcome information of 57 patients with positive RT-PCR assay for COVID-19.

Two experienced thoracic subspecialty radiologists evaluated all chest CT examinations and recorded opacity type, distribution and extent of lobar involvement. Information on symptom duration prior to hospital admission, duration of hospital admission, presence of comorbid conditions, laboratory data, and outcomes (recovery or death) was obtained from the medical records. Entire lung volume was segmented on thin-section DICOM images (1.5-2 mm) to obtain whole-lung analysis. Statistics of the datasets are shown in Tables~\ref{tab:feature_statistics}-\ref{tab:feature_novara} in Section~\ref{sec:DVB}.

\section{ICU Admission Prediction}

\begin{figure}[t]
\includegraphics[width=\textwidth]{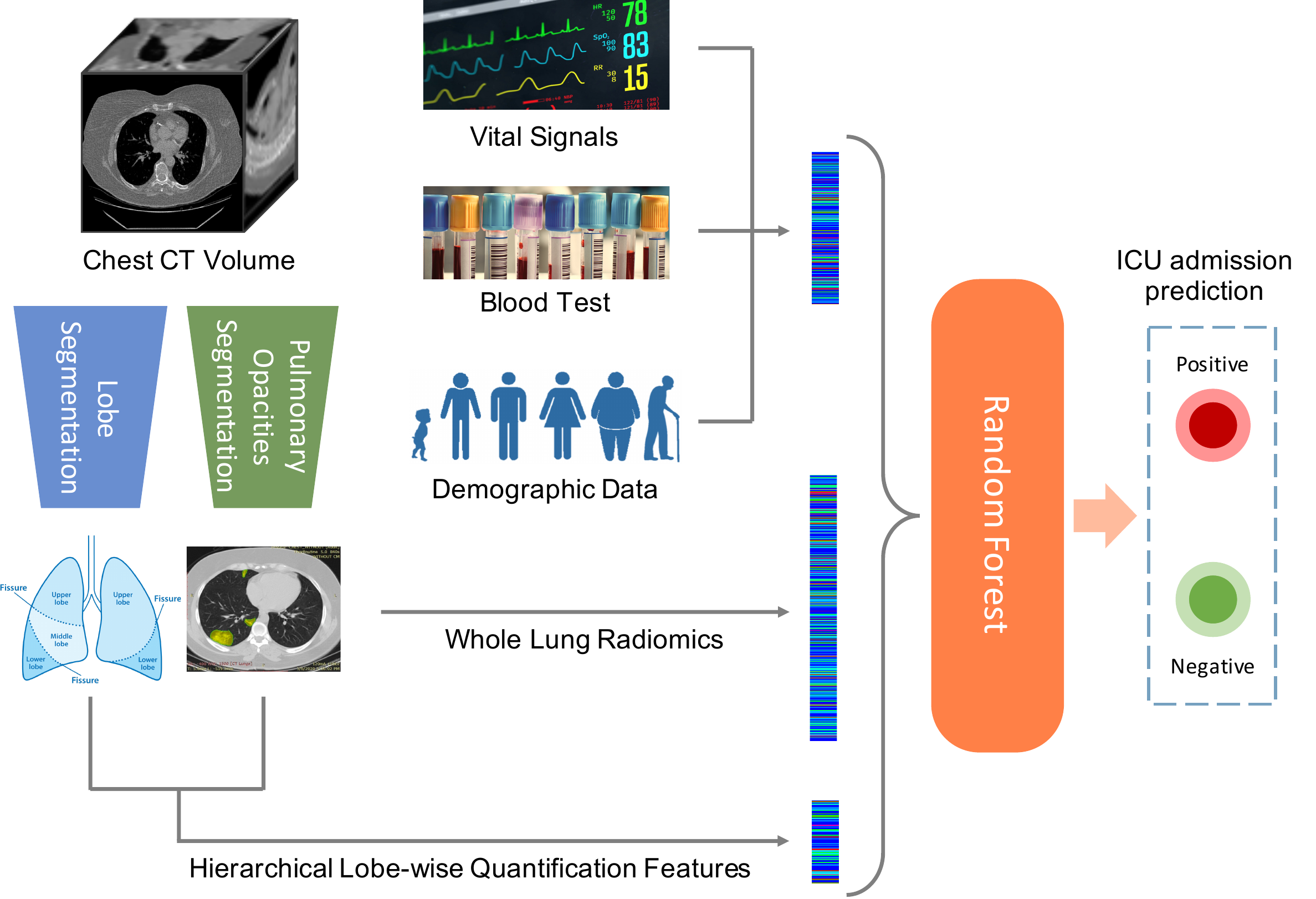}
\caption{Framework of the proposed methods including the utilized inputs and expected output.}
\label{fig:overview}
\end{figure}

In order to predict the need for ICU admission of patients with COVID-19 pneumonia, we use three types of imaging and non-imaging features. Our adopted features include hierarchical lobe-wise quantification features (HLQ), whole lung radiomics features (WLR), and features from demographic, vital signs, and blood examination (DVB). Figure~\ref{fig:overview} shows an overview of the overall framework of the presented work. In the rest of this section, we first introduce the details of these features. Since it is challenging to fuse the large number of inhomogeneous features together, a feature selection strategy is proposed, followed by random forest based classification~\citep{breiman2001random}.

\subsection{Deep Learning based Image Segmentation}\label{section:deep}

In our work, we employed deep neural networks to segment both lungs, five lung lobes and pulmonary opacities (as regions of infection) from non-contrast chest CT examinations. For training purpose, we semi-automatically labeled 71 CT volumes using 3D Slicer \citep{Kikinis2014}.
For lung lobe segmentation, we adopted the automated lung segmentation method by \cite{hofmanninger2020automatic}. 
The pre-trained model\footnote{https://github.com/JoHof/lungmask} was fine-tuned with a learning rate of $1\times 10^{-5}$  using our annotated data. 
The tuned model was then applied to segment all the chest CT volumes.

Segmentation of pulmonary opacities was completed by our previously proposed method, Pyramid Input Pyramid Output Feature Abstraction Network (PIPO-FAN)~\citep{fang2020multi} with publicly released source code\footnote{https://github.com/DIAL-RPI/PIPO-FAN}.

\begin{figure}
\centering
\subfloat[COVID-19 CT image]
{\includegraphics[width=.33\textwidth]{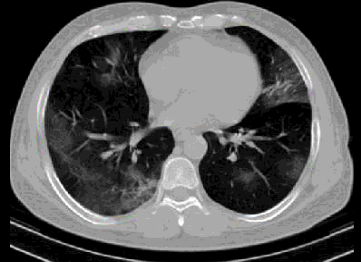}}
\hfill
\subfloat[Lobes and infection]
{\includegraphics[width=.33\textwidth]{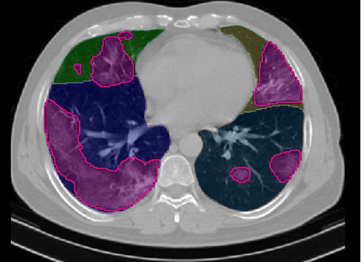}}
\hfill
\subfloat[3D Visualization]
{\includegraphics[width=.32\textwidth]{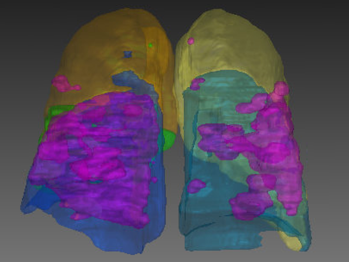}}
\caption{Lung lobes and pulmonary opacities segmentation results. Areas colored in magenta indicate the segmented lesions.}\label{fig:segmentation}
\end{figure}

Figure~\ref{fig:segmentation} shows the segmentation results of lung lobes and pulmonary opacities. From axial and 3D view, we can see that the segmentation models can smoothly and accurately predict isolate regions with pulmonary opacities.

\subsection{Hierarchical Lobe-wise Quantification Features}
\label{Sec:HLQF}

Based on the segmentation results in Section~\ref{section:deep}, we then compute the ratio of opacity volume over different lung regions, which is a widely used measurement to describe the severity \citep{tang_severity_2020, zhu_joint_2020}. The lung regions include the whole lung, the left lung and the right lung and 5 lung lobes (lobe\# 1-5) as shown in Figure~\ref{fig:segmentation}. The right lung includes upper lobe (lobe\# 1) , middle lobe (lobe\# 2), and lower lobe (lobe\# 3), while left lung includes upper lobe(lobe\# 4) and lower lobe (lobe\# 5). Thus, each CT image has 8 regions of interest (ROIs). In addition to the ROI segmentation, we partitioned each segment to 4 parts based on the HU ranges, i.e., the HU ranges of -$\infty$ to -750 (HU[-$\infty$, -750]),  -750 to -300 (HU[-750, -300]),  -300 to 50 (HU[-300, 50]), and 50 to $+\infty$ (HU[50, $+\infty$]). These four HU ranges correspond to normal lungs, ground glass opacity (GGO), consolidation, and regions with pulmonary calcification, respectively. As a result, each CT image was partitioned to 32 components (8 ROIs $\times$ 4 ranges/ROI).

We extracted two quantitative features from each part, i.e., volumes of pulmonary opacities (VPO) and ratio of pulmonary opacities to the corresponding component (RPO), as defined below:
\begin{equation}
VPO(x) = V(Segment(x))\\
\end{equation}
\begin{equation}
RPO(x) = \frac{VPO(x)}{V(x)} = \frac{V(Segment(x))}{V(x)}\\
\end{equation}
where $x$ is a selected component (among the 32 components). $Segment(x)$ denotes the pulmonary opacities in the selected component $x$ based on the segmentation of pulmonary opacities in Figure~\ref{fig:segmentation}. $V(\cdot)$ denotes the volume of the selected part.

\subsection{Whole Lung Radiomics Features}

\begin{table}[t]
	\centering
	\caption{Radiomics feature types and number of each kind of features.}\label{tab:featuretype}
	\begin{tabular}{|l|l|c|c|}
		\hline
		\textbf{Group} & \textbf{Feature type} & \textbf{\# features} & \textbf{Sum}\\
		\hline
		\multirow{6}{*}{Texture} & First order & 18 & \multirow{6}{*}{93}\\
		\cline{2-3}
		& GLCM & 24 & \\
		\cline{2-3}
		& GLRLM & 16 & \\
		\cline{2-3}
		& GLSZM & 16 & \\
		\cline{2-3}
		& NGTDM & 5 & \\
		\cline{2-3}
		& GLDM & 14 & \\
		\hline
		Shape & Shape (3D) & 17 & 17\\
		\hline
	\end{tabular}
\end{table}
To more comprehensively describe information in the CT image, we also extracted multi-dimensional radiomics features~\citep{gillies2015} of all pulmonary opacities. Compared with HLQ feature, although they are all image-based features, they describe the pulmonary opacities from different aspects. HLQ features focus on the pulmonary opacities volume and position of region of interest, while WLR focus on their shape and texture.

For each chest CT volume, we first masked out non-infection regions based on the infection segmentation results, then four kinds of radiomics features are calculated on the volume, \textit{i.e.}, shape, first-order, second-order and higher-order statistics features~\citep{rizzo2018}. Shape features describe the geometric information.  First-order, second-order and higher-order statistics features all describes texture information. First-order statistics features describe the distribution of individual voxel values without concerning spatial correlation. Second-order features describe local structures, which provide correlated information between adjacent voxels and statistical measurement of intra-lesion heterogeneity. Second-order features include those extracted using gray level dependence matrix (GLDM), gray level co-occurrence matrix (GLCM), grey level run length matrix (GLRLM), grey level size zone matrix (GLSZM), and neighboring gray tone difference matrix (NGTDM). Higher-order statistics features are computed using the same methods as second-order features but after applying wavelets and Laplacian transform of Gaussian(LoG) filters. The higher-order features help identify repetitive patterns in local spatial-frequency domain in addition to suppressing noise and highlighting details of images.

\begin{table}[t]
\centering
\caption{Image filter types and extracted radiomics features from each type of filtered images.}
\label{tab:radiomics}
\begin{tabular}{|l|l|l|}
\hline
\textbf{Image filter type} & \textbf{Extracted features} & \textbf{\# features}\\
\hline
No filter (Original image) & Texture + Shape & 93+17=110\\
\hline
Square filter & Texture & 93\\
\hline
Square-root(Sqrt) filter & Texture & 93\\
\hline
Logarithm filter & Texture & 93\\
\hline
Exponential filter & Texture & 93\\
\hline
Wavelet filters & \multirow{3}{*}{Texture} &  \multirow{3}{*}{93$\times$8=744}\\
(HHH, HHL, HLH, LHH, & & \\
HLL, LHL, LLH, LLL) & &\\
\hline
Laplacian of Gaussian (LoG) filters & \multirow{2}{*}{Texture} &  \multirow{2}{*}{93$\times$5=465}\\
$\sigma$ $\in$ \{0.5, 1.5, 2.5, 3.5, 4.5\} & &\\
\hline

\end{tabular}
\end{table}

We used the Pyradiomics package \citep{griethuysen2017} to extract the above described radiomics features from COVID19 chest CT images. For each chest CT volume, a total of 1691 features are extracted. The number of radiomics features for each feature type are summarized in Table~\ref{tab:featuretype}. Based on the description above, these features can be categorized into two main groups, \emph{i.e.}, 17 shape features and 93 texture features.

To extracted various features, different image filters are applied before feature extraction. Table~\ref{tab:radiomics} shows the details of all 18 image filter types used in our work, including no filter, square filter, square-root filter, logarithm filter, exponential filter, wavelet filter and LoG filter. The image filtered by a 3D wavelet filter has eight channels, including HHH, HHL, HLH, LHH, HLL, LHL, LLH and LLL. The Laplacian of Gaussian (LoG) filters have a hyper-parameter $\sigma$ which is the standard deviation of the Gaussian distribution. We used five different $\sigma$ values in our study, \textit{i.e.}, $\{0.5, 1.5, 2.5, 3.5, 4.5\}$. Note that shape features are only extracted from the original images (no filter was applied).

\subsection{Non-imaging Features}
\label{sec:DVB}
\begin{table}[hbt]
\caption{Statistics (mean$\pm$std, except for gender) of DVB features for Site A dataset.}
\centering
\scalebox{0.9}{
\begin{tabular}{|l|rcl|rcl|c|}
\hline
\textbf{ICU admission} & \multicolumn{3}{c|}{\textbf{Not Admitted}} & \multicolumn{3}{c|}{\textbf{ICU Admitted}} & \textbf{Data \#}\\
\hline
Gender (M:F) & $43$ & $:$ & $28$ & $29$ & $:$ & $13$ & 113\\
Age (year) & $56.7$ & $\pm$ & $16.0$ & $66.9$ & $\pm$ & $16.2$ & 113\\
\hline
Lym\_r (\%) & $22.7$ & $\pm$ &  $8.3$ & $15.6$ & $\pm$ & $12.8$ & 113\\
WBC  & $5831.0$ & $\pm$ & $1848.9$ & $7966.7$ & $\pm$ & $4556.2$ & 113\\
Lym  & $1244.7$ & $\pm$ & $482.8$ & $1010.4$ & $\pm$ & $943.7$ & 113\\
\hline
Temperature (\textcelsius{}) & $37.3$ & $\pm$ & $0.6$ & $37.6$ & $\pm$ & $0.6$ & 98\\
SpO2 (\%) & $91.9$ & $\pm$ & $7.41$ & $86.5$ & $\pm$ & $8.53$ & 100\\
\hline
\end{tabular}}
\label{tab:feature_statistics}
\end{table}

\begin{table}[hbt]
\caption{Statistics (mean$\pm$std, except for gender) of DVB features for Site B dataset.}
\centering
\scalebox{0.9}{
\begin{tabular}{|l|rcl|rcl|c|}
\hline
\textbf{ICU admission} & \multicolumn{3}{c|}{\textbf{Not Admitted}} & \multicolumn{3}{c|}{\textbf{ICU Admitted}} & \textbf{Data \#}\\
\hline
Gender (M:F) & $23$ & $:$ & $24$ & $39$ & $:$ & $39$ & 125\\
Age (year) & $74.8$ & $\pm$ & $15.0$ & $72.7$ & $\pm$ & $11.1$ & 125\\
\hline
Lym\_r (\%) & $18.6$ & $\pm$ &  $12.7$ & $13.0$ & $\pm$ & $12.8$ & 125\\
WBC  & $7175.7$ & $\pm$ & $4288.9$ & $11722.3$ & $\pm$ & $7249.3$ & 125\\
Lym  & $1058.1$ & $\pm$ & $596.7$ & $1613.8$ & $\pm$ & $3872.7$ & 125\\
\hline
\end{tabular}}
\label{tab:feature_mgh}
\end{table}

\begin{table}[hbt]
\caption{Statistics (mean$\pm$std, except for gender) of DVB features for Site C dataset.}
\centering
\scalebox{0.9}{
\begin{tabular}{|l|rcl|rcl|c|}
\hline
\textbf{ICU admission} & \multicolumn{3}{c|}{\textbf{Not Admitted}} & \multicolumn{3}{c|}{\textbf{ICU Admitted}} & \textbf{Data \#}\\
\hline
Gender (M:F) & $13$ & $:$ & $8$ & $24$ & $:$ & $12$ & 57\\
Age (year) & $70.0$ & $\pm$ & $13.7$ & $66.9$ & $\pm$ & $12.3$ & 57\\
\hline
Temperature (\textcelsius{}) & $39.0$ & $\pm$ & $1.0$ & $37.8$ & $\pm$ & $0.9$ & 50\\
SpO2 (\%) & $92.3$ & $\pm$ & $5.25$ & $84.5$ & $\pm$ & $7.74$ & 31\\
\hline
\end{tabular}}
\label{tab:feature_novara}
\end{table}

In addition to features extracted from images, we incorporated features from demographic data (contained by all three datasets), vital  signs (from Sites A and B),  and laboratory data (from  Sites A and C) (DVB).  Specifically, such features include patients' age, gender, white blood cell count (WBC), lymphocyte count (Lym), Lym to WBC ratio (L/W ratio), temperature and blood oxygen level (SpO2). These data are highly correlated with the ICU admission of patients when they were admitted to a hospital. Table~\ref{tab:feature_statistics}-\ref{tab:feature_novara} show the statistics of the above features in Site A, Site B, and Site C datasets respectively. Non-imaging features are not all available for some patients. The number of the collected data for each feature is listed in the last column of the tables. To make use of all the data, the missing values are imputed by the mean values of other available entries.  For instance, in Site A dataset, if a patient’s SpO2 was not recorded, the mean SpO2 value of 91.9 from the dataset is used to fill the blank.

\subsection{ICU Admission Prediction}
\label{sec:prediction}

In our work, random forest (RF) \citep{breiman2001random} classifier, a widely-used ensemble learning method consisting of multiple decision trees, is chosen for predicting ICU admission due to its several nice properties. First, RF is robust to small data size. Second, it can generate feature importance ranking and is thus highly interpretable. Aggregating all the features introduced above, we have 1,762 features in total. Due to the limited data size, the model would easily overfit with all features as input. Thus, we first used RF to rank all the features, then we selected the top $K$ features for our final prediction.

We ranked the feature based on their Gini importance \citep{leo1984classification}. It is calculated during the training of RF by averaging the decrease of Gini impurity over all trees. Due to the randomness of RF, Gini importance of features may vary when RF model is initialized with different random seeds. Therefore, in our study, feature ranks are computed $100$ times with different random seeds. Each time every feature will get a score being its rank. The final feature rank is obtained by sorting the total summed score of each feature.

Based on the rank of all the features, we select top $K\in$[1,100] features to train the RF model and calculate the prediction performance in terms of AUC.

\section{Experimental Results}

This section presents the experimental results of the developed methods. We show the effectiveness of our proposed method on the three datasets separately through both ablation studies and comparison with other state-of-the-art approaches. We did not merge the datasets because of two reasons. First, not all the non-image features were available from the participating sites.  Second, the treatment and admission criteria at the participating sites were likely different from each other. Given such limitations, the datasets were used separately to evaluate the proposed methods.

\begin{figure}[t]
	\centering
		\includegraphics[width=\textwidth]{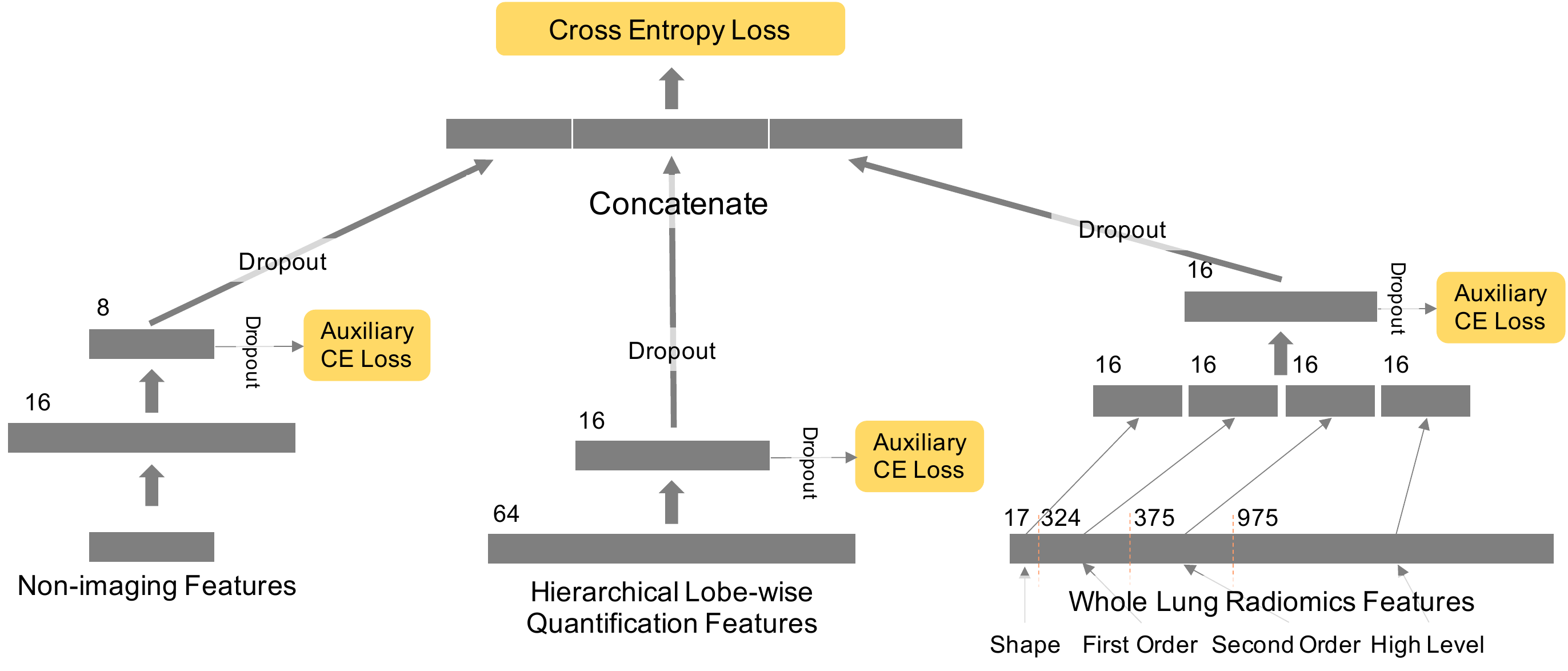}	
	\caption{Architecture of the Wide \& Deep Net~\citep{cheng2016wide} based deep neural network (DNN). Three different kinds of features are first processed separately by one or two fully connected layers. Then the learned features are concatenated for the final prediction.}
	\label{fig:net}
\end{figure}

The experiments are summarized in to two parts. In the first part, the proposed methods with different combinations of features is compared with other state-of-the-art approaches on each dataset. 
In this part of the experiments, we also included results of support vector machine (SVM), logistic regression and three other deep learning networks. The first deep neural network (DNN) takes all the WLR, HLQ and DVB features as its input and consists of three fully connected layers with output dimensions of 64, 16 and 2 respectively (denoted as \emph{DNN w/ all features}). Dropout was applied on the first two layers with $50\%$ dropping rate. The second deep network takes features selected by random forest as its input. As the selected features is only a small subset of all the features, this network is smaller than the first one. It contains three fully connected layers with output dimension of 8, 8 and 2, respectively (denoted as \emph{DNN (small)}). The third network is designed based on the Wide \& Deep Net (WD Net)~\citep{cheng2016wide} to investigate whether a more complex network could perform better directly using all features as an input (denoted as WD Net w/ all features). The detailed structure is shown in Fig.~\ref{fig:net}. For all the three networks, the cross entropy loss is used for training. For a sample with label $y\in\{0,1\}$, the cross entropy loss is formulated as $l=-log P_y$, where $P_y$ is the output prediction probability of class $y$. In the five-fold cross validation scheme, each time 3 folds are used to train the network, one fold is used as validation set, and the last one is reserved as test set. The networks were implemented over PyTorch~\citep{NEURIPS2019_9015} and trained using the Adam optimizer with learning rate of $1e-4$. Influenced by the size of dataset and the network, the WD Net on Site B dataset took the longest time for training which is $3.0$ min with 4 NVIDIA Tesla V100 GPUs. In the second part of the experiments, the generalization ability of the feature combination learned by our model is studied across three datasets. The code is open sourced and available at  \url{https://github.com/DIAL-RPI/COVID19-ICUPrediction}.

Several recent works have shown the importance of using machine learning models to predict patients' outcomes based on lobe-wise quantification features. The infection volume and infection ratio of the whole lung, right/left lung, and each lobe/segment are calculated as quantitative features in \cite{tang_severity_2020}. Random forest classifier is used to select the top-ranking features and make the severity assessment based on these features. In another work by \cite{zhu_joint_2020}, the authors present a novel joint regression and classification method to identify the severity cases and predict the conversion time from a non-severe case to the severe case. Their lobe-wised quantification features include the infection volume, density feature and mass feature. As we mentioned earlier, all existing image analysis-based outcome prediction works use only image features. We take the features in the two papers as baseline to compare with our work.

\subsection{Results on Site A Dataset}
\label{sec:exp_iran}
\begin{table}[hbt]
\centering
\caption{Comparison among the features used in exist state-of-the-art works and different combinations of the proposed features on ICU admission prediction on Site A dataset. One-tailed $t$-test is used to evaluate the statistical significance between a feature combination and the best performer.}
\label{tab:iran_res}
\scalebox{0.72}{
\begin{tabular}{|l|c|c|c|c|c|c|c|}
\hline
\multirow{2}{*}{\textbf{Features}} & \multicolumn{3}{c|}{\textbf{AUC}} &  \multicolumn{3}{c|}{\textbf{Sensitivity (PPV$=$70\%)}} & \multirow{2}{*}{\textbf{K}} \\
\cline{2-7}
& Mean & 95\% CI & p Value & Mean & 95\% CI & p Value &\\
\hline
Img feature (Tang 2020)  & 0.818 & (0.796, 0.839) & $p<.001$ & 51.0\% & (39.3\%, 62.6\%) & $p<.001$ & 8 \\
\hline
Img feature (Zhu 2020) & 0.776 & (0.762, 0.790) & $p<.001$ & 48.6\% & (35.4\%, 61.7\%) & $p=.001$ & 46\\
\hline
\hline
DVB & 0.855 & (0.844, 0.866) & $p=.002$ & 76.7\% & (73.2\%, 80.1\%) & $p=.017 $ & 1\\
\hline
HLQ & 0.789 & (0.781, 0.797) & $p<.001$ & 51.4\% & (45.3\%, 57.5\%) & $p<.001$  & 21\\
\hline
WLR & 0.859 & (0.843, 0.873) & $p<.001$ & 71.4\% & (60.5\%, 82.3\%) & $p=.022 $ & 70\\
\hline
WLR+HLQ & 0.866 & (0.857, 0.875) & $p<.001$ & 68.6\% & (57.6\%, 79.5\%) & $p=.008$  & 61\\
\hline
WLR+DVB & \emph{0.876} & (0.867, 0.886) & \emph{p=.109} & \emph{81.4\%} & (76.0\%, 86.8\%) & \emph{p=.152} & 4\\
\hline
HLQ+DVB & \emph{0.865} & (0.844, 0.885) & \emph{p=.080} & 70.0\%  & (60.9\%, 79.1\%) & $p=.012$ & 4\\
\hline
WLR+HLQ+DVB & \textbf{0.884} & (0.875, 0.893) & - & \textbf{84.3\%} & (79.9\%, 88.7\%) & - & 52\\
\hline
\end{tabular}
}
\end{table}

\begin{figure}[hbt]
	\centering
	\includegraphics[width=1\textwidth,trim=50 20 70 50, clip]{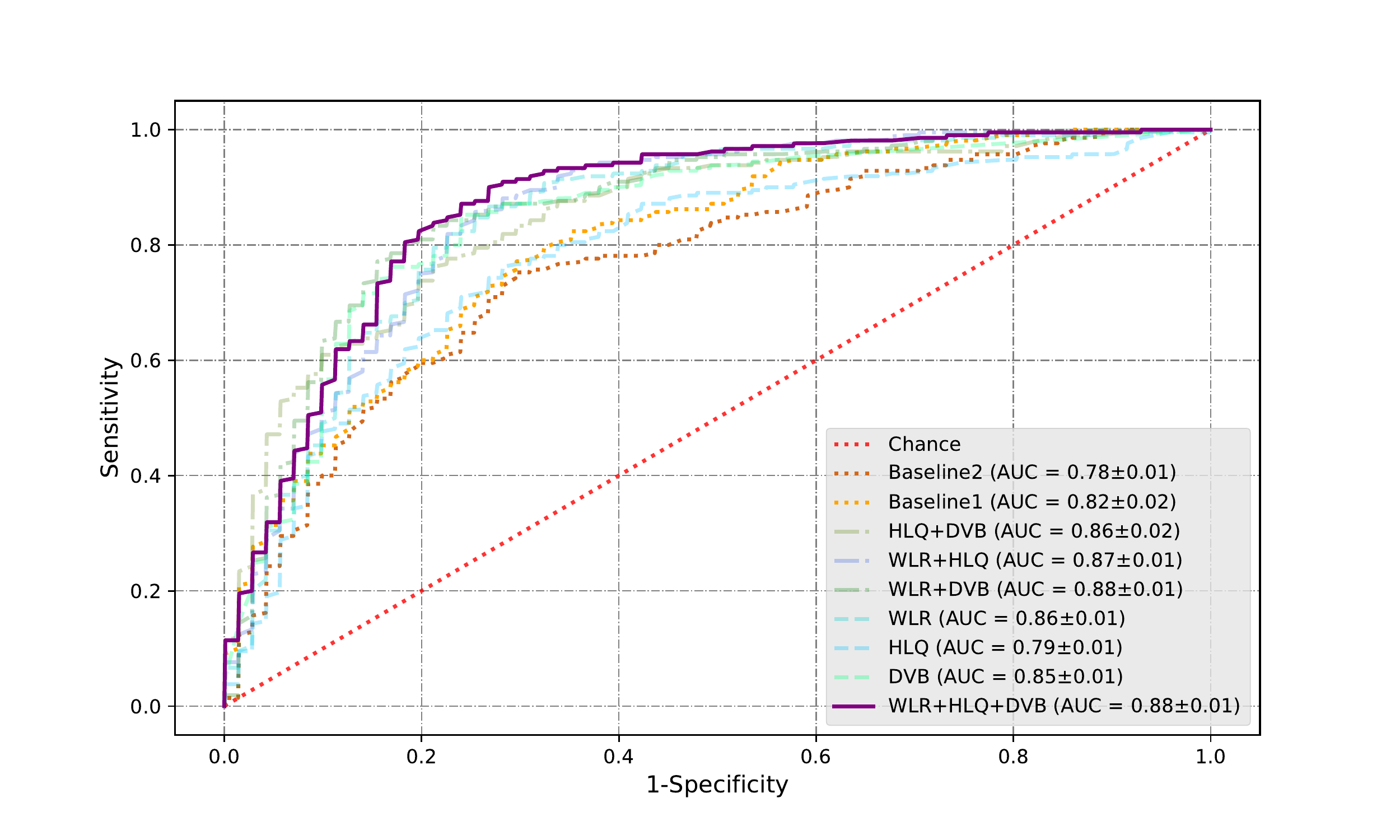}
	
\caption{ROC curves of various feature combinations on Site A dataset. \textbf{DVB}: non-imaging features including Demographic data, Vital signals and Blood test results; \textbf{HLQ}: Hierarchical Lobe-wise Quantification features; \textbf{WLR}: Whole Lung Radiomics features.}
\label{fig:iran_auc}
\end{figure}

\begin{figure}[h!]
	\centering
	\includegraphics[width=.8\textwidth,trim={12 10 10 10},clip]{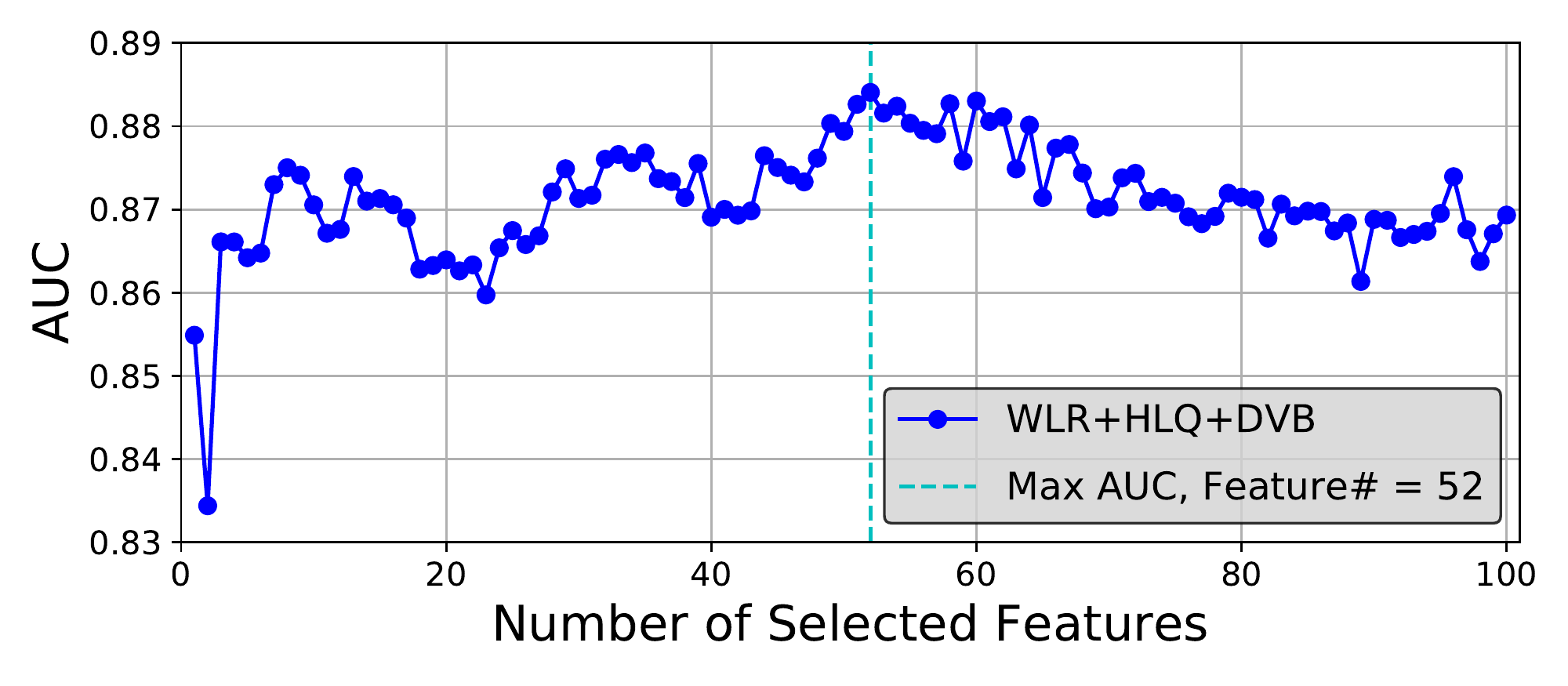}
	\caption{Variation of AUC along choosing the top $K$ features.}
	\label{fig:K-feature_selection}
\end{figure}

Receiver Operating Characteristics (ROC) curves of the feature combinations are shown in Figure~\ref{fig:iran_auc}. For each feature combination, the features are selected only from the feature categories available in the combination using the approach introduced in Section~\ref{sec:prediction}. For example, HLQ+DVB indicates that only features from these two groups, HLQ and DVB, are selected and used. The number of features $K$ used to obtain the best results are listed in Table~\ref{tab:iran_res}. To alleviate the stochasticity of the results, for each feature combination, five RF models with different random seeds are trained and tested with five fold cross validation. The curves shown here are thus the mean results of the five models. The figure legend gives the mean Area Under the Curves (AUCs) of the feature combinations as well as the standard deviation (mean$\pm$std). It can be seen that the combination of all three kinds of features, WLR+HLQ+DVB, obtained the best result with an AUC of 0.88$\pm$0.01. The variation of AUC along with number of selected features on Site A dataset is presented in Fig.~\ref{fig:K-feature_selection}. As marked by the light blue dash line in Fig.~\ref{fig:K-feature_selection}, AUC reaches the maximum value when the top 52 features are selected. Details of the 52 selected features are presented in Table~\ref{tab:feature_rank_siteA} at the end of this paper due to its large size.

One-tailed $t$-test is used to evaluate the statistical significance between a method and the best performer.
Table~\ref{tab:iran_res} summarizes the AUC values and sensitivity with significance test $p$ values and 95\% confidence interval (95\% CI). The classification threshold is selected by control the positive prediction value (PPV) to be 70\%. The combination of WLR+HLQ+DVB significantly exceeds other reported methods~\citep{tang_severity_2020, zhu_joint_2020} with $p\le0.001$. Further, we achieved a sensitivity of 84.3\% while retaining a PPV at 70\%. It suggests that our model can rapidly prioritize over 80\% patients who would develop into critical conditions, if we allow 3 false positive cases in every 10 positive predictions. With such prediction, hospitals may allocate limited medical resources more efficiently to potentially prevent such conversion and save more lives. Under the same setting, the sensitivity of the model is 79.4\%, the accuracy is 81.2\%.

\begin{table}[hbt]
\centering
\caption{Comparison of different machine learning methods with selected features on Site A dataset. One-tailed $t$-test is used to evaluate the statistical significance between a feature combination and the best performer.}
\label{tab:iran_mth}
\scalebox{0.72}{
\begin{tabular}{|l|c|c|c|c|c|c|}
\hline
\multirow{2}{*}{\textbf{Methods}} & \multicolumn{3}{c|}{\textbf{AUC}} &  \multicolumn{3}{c|}{\textbf{Sensitivity (PPV$=$70\%)}} \\
\cline{2-7}
& Mean & 95\% CI & p Value & Mean & 95\% CI & p Value\\
\hline
Random Forests & \textbf{0.884} & (0.875, 0.893) & - & \textbf{84.3\%} & (79.9\%, 88.7\%) & - \\
\hline
SVM & 0.867 & (0.855, 0.880) & $p=.002$ & 71.0\% & (64.9\% , 77.0\%) & $p<.001$ \\
\hline
Logistic Regression & 0.785 & (0.758, 0.812) & $p<.001$ & 31.0\% & (14.8\%, 47.1\%) & $p<.001$\\
\hline
\hline
DNN (small) & 0.816 & (0.804, 0.828) & $p<.001$ & - & - & $p=0.023$\\
\hline
DNN w/ all features & 0.751 & (0.723, 0.779) & $p<.001$ & 25.7\% & (0.8\%, 50.6\%) & $p=.003$\\
\hline
WD Net w/ all features & 0.823 & (0.807, 0.838) & $p<.001$ & 58.1\% & (40.7\%, 75.5\%) & $p=.009$\\
\hline
\end{tabular}
}
\end{table}

In the comparison among different combinations of the features, we can see that the results are generally improved with more feature sources added. Comparison between WLR+HLQ (line 6) and WLR+HLQ+DVB (the last line) shows that, on this dataset, introducing non-imaging features can significantly improve the performance (p$<$0.001 for AUC and p=0.008$<$0.05 for sensitivity), which further indicates that non-imaging features and image based features are complementary. On the other hand, the comparison with WLR+DVB (line 7) and HLQ+DVB (line 8) shows that the improvement of WLR+HLQ+DVB was not significant (p$>$0.05). It suggests that different kinds of image based features may contain redundant information and adding more features from the same source only results in marginal improvement. Table~\ref{tab:iran_mth} shows the results of different methods on Site A dataset. Unless specially noted as “w/ all features”, the methods in Table~\ref{tab:iran_mth} use 52 selected WLR+HLQ+DVB features listed in Table~\ref{tab:feature_rank_siteA}. We can see that random forest performed significantly better ($p<0.05$) than all the other methods on both AUC and sensitivity. The sensitivity of DNN (small) is not included because it couldn’t obtain a PPV equal or larger than $70\%$ in some of the cross validation folds.

\subsection{Results on Site B Dataset}

\begin{table}[hbt]
\centering
\caption{Comparison among the features used in exist state-of-the-art works and different combinations of the proposed features on ICU admission prediction on Site B dataset.}
\label{tab:mgh_res}
\scalebox{0.72}{
\begin{tabular}{|l|c|c|c|c|c|c|c|}
\hline
\multirow{2}{*}{\textbf{Features}} & \multicolumn{3}{c|}{\textbf{AUC}} &  \multicolumn{3}{c|}{\textbf{Sensitivity (PPV$=$70\%)}} & \multirow{2}{*}{\textbf{K}} \\
\cline{2-7}
& Mean & 95\% CI & p Value & Mean & 95\% CI & p Value &\\
\hline
Img feature (Tang 2020) & 0.770 & (0.745, 0.796) & $p<.001$ & 83.1\% & (75.8\%, 90.4\%) & $p=.009$ & 10\\
\hline
Img feature (Zhu 2020) & 0.767 & (0.752, 0.781) & $p<.001$ & 83.8\% & (82.2\%, 85.5\%) & $p<.001$ & 39\\
\hline
\hline
DVB & 0.671 & (0.643, 0.700) & $p<.001$ & 78.7\% & (69.7\%, 87.7\%) & $p=.007 $ &  4\\
\hline
HLQ & 0.791 & (0.774, 0.809) & $p<.001$ & 84.6\% & (81.3\%, 88.0\%) & $p<.001$  &  3\\
\hline
WLR & 0.841 & (0.827, 0.855) &$p=.014$ & \textbf{94.9\%} & (93.4\%, 96.3\%) & - & 55\\
\hline
WLR+HLQ & \textbf{0.847} & (0.833, 0.861) & - & \emph{92.6\%} & (89.5\%, 95.7\%) & $p=.083$  & 12\\
\hline
WLR+DVB & 0.841 & (0.828, 0.854) & $p=.257$ & 91.8\% & (90.5\%, 93.1\%) & $p=.012$ & 33\\
\hline
HLQ+DVB & 0.796 & (0.777, 0.815) & $p=.001$ & 84.4\%  & (80.7\%, 88.0\%) & $p<.001$ & 4\\
\hline
WLR+HLQ+DVB & \emph{0.844} & (0.833, 0.855) & $p=.310$ & \emph{92.6\%} & (90.0\%, 95.1\%) & $p=.027$ & 12\\
\hline

\end{tabular}
}
\end{table}

\begin{figure}[h!]
	\centering
	\includegraphics[width=1\textwidth,trim=50 20 70 50, clip]{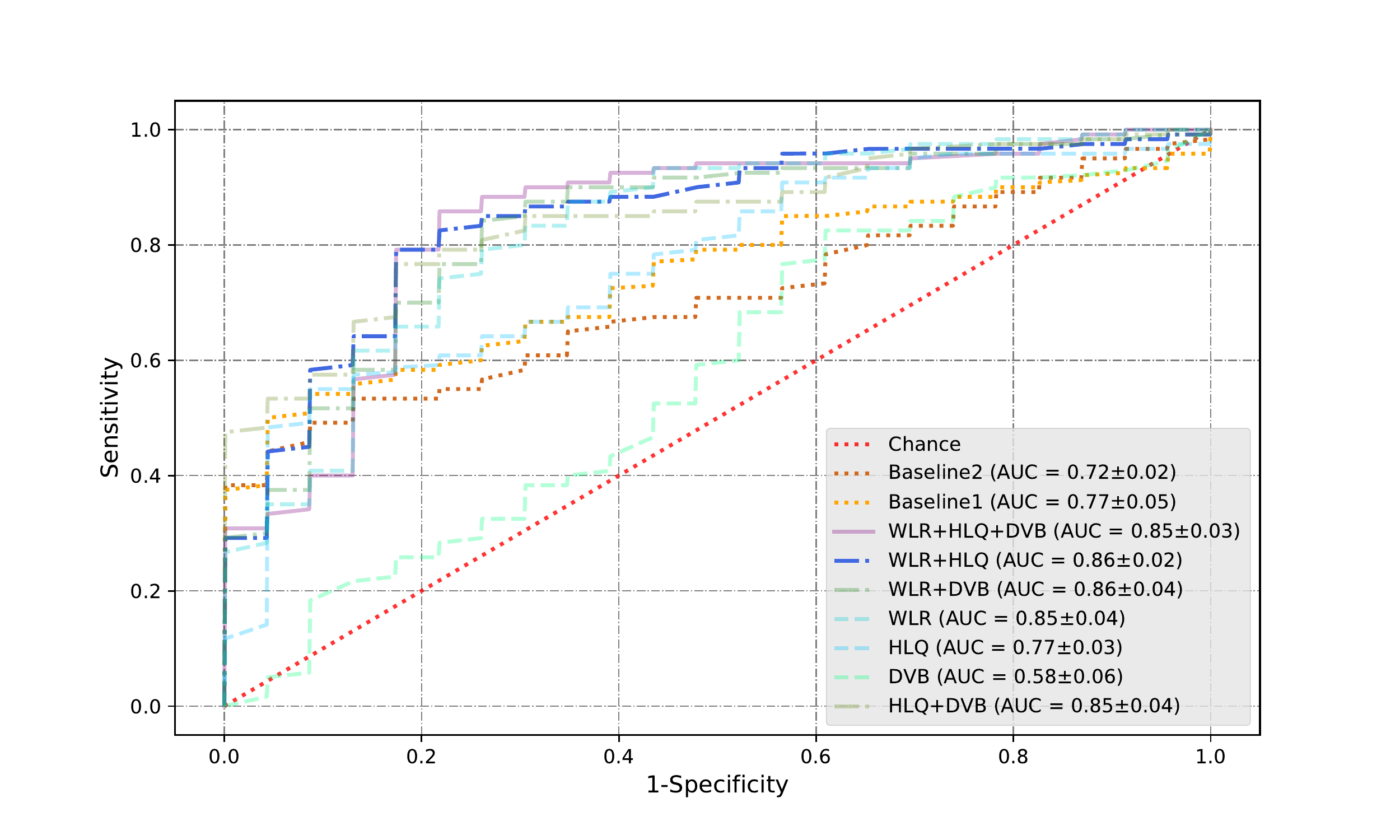}
	
	\caption{ROC curves on Site B dataset.}\label{fig:mgh_auc}
\end{figure}

\begin{table}[h!]
\centering
\caption{The top 12 WLR+HLQ features ranked by feature ranking strategy introduced in Section~\ref{sec:prediction} on Site B dataset. The third and sixth columns show the Gini importance of the corresponding feature averaged in the 5-fold cross validation.}
\label{tab:feature_rank_siteB}
\scalebox{0.7}{
\begin{tabular}{|l|l|l||l|l|l|}
\hline
\textbf{\#} & \textbf{Quantitative features} & \textbf{G(\%)} & \textbf{\#} & \textbf{Quantitative features} & \textbf{G(\%)} \\
\hline
    1 & {\color{ForestGreen}Lobe\#2 RPO HU3\tnote{*}} & 10.27 & 2 & {\color{ForestGreen}Lobe\#2 RPO HU2}& 8.45 \\
    \hline
    3 & {\color{ForestGreen}Whole Lung RPO HU1} & 10.77 & 4 & {\color{blue}LoG($\sigma=4.5$)-firstorder-Kurtosis\tnote{**}}& 7.74 \\
    \hline
    5  & {\color{blue} Exponential-GLRLM-ShortRunLowGL}& 7.8 & 6 & {\color{blue}LoG($\sigma=4.5$)-NGTDM-Contrast}& 8.07 \\
    \hline
    7 & {\color{blue}Exponential-GLRLM-ShortRun} & 7.24 & 8 & {\color{blue}LoG($\sigma=3.5$)-GLSZM-ZoneEntropy}& 7.58 \\
    \hline
    9 & {\color{blue}LoG($\sigma=4.5$)-NGTDM-Busyness}& 7.0 & 10 & {\color{blue}Sqrt-NGTDM-Strength}& 9.0 \\
    \hline
    11 & {\color{blue}LoG($\sigma=3.5$)-GLCM-Imc2}& 8.17 & 12 & {\color{blue}LLH-GLSZM-GLNonUnif}& 7.89 \\
    \hline
\end{tabular}
}
\begin{tablenotes}
	\footnotesize
	\item[{\color{ForestGreen}*}] {\color{ForestGreen}Green text} indicates lobe-wise quantification features, HU1-HU4 are the four HU intervals.
	\item[\blue{**}] \blue{Blue text} indicates whole lung radiomics features encoded as Filter-FeatureType-Parameter.
\end{tablenotes}
\end{table}

The same set of experiments were repeated on the Site B dataset. Table~\ref{tab:mgh_res} and Fig.~\ref{fig:mgh_auc} shows the results. The number of features $K$ used to obtain the best results for each combination are listed in Table~\ref{tab:mgh_res}. It can be seen that, on Site B dataset, non-imaging features are not very predictive. There could be several reasons for the inability of DVB features on Site B dataset. First, as shown in Table.~\ref{tab:feature_mgh} non-imaging features on Site B dataset have a large standard deviation. Second, the use of CT in Site B is different from that in Site A. Site B relied on chest radiography for most patients while CT was reserved for more sicker patients or those with suspected complications; Site A used CT in all patients regardless of clinical  severity. Third, criteria of ICU admission are different between two sites. Fourth, management strategies and disease outcomes at the two sites are different.

\begin{table}[hbt]
\centering
\caption{Comparison of different machine learning methods with selected features on Site B dataset.}
\label{tab:mgh_mth}
\scalebox{0.72}{
\begin{tabular}{|l|c|c|c|c|c|c|}
\hline
\multirow{2}{*}{\textbf{Methods}} & \multicolumn{3}{c|}{\textbf{AUC}} &  \multicolumn{3}{c|}{\textbf{Sensitivity (PPV$=$70\%)}} \\
\cline{2-7}
& Mean & 95\% CI & p Value & Mean & 95\% CI & p Value\\
\hline
Random Forests & 0.844 & (0.833, 0.855) & - & 92.6\% & (90.0\%, 95.1\%) & - \\
\hline
SVM & \textbf{0.852} & (0.838, 0.866) & $p=.148$ & \textbf{95.4\%} & (93.2\% , 97.5\%) & $p=.020$ \\
\hline
Logistic Regression & 0.798 & (0.783, 0.812) & $p=.003$ & 90.2\% & (88.6\%, 91.9\%) & $p=.110$\\
\hline
\hline
DNN (small) & 0.831 & (0.805, 0.858) & $p=.187$ & 86.9\% & (83.5\%, 90.3\%) & $p=.003$\\
\hline
DNN w/ all features & 0.704 & (0.670, 0.738) & $p<.001$  & 75.38\% & (71.6\%, 79.2\%) & $p=.001$\\
\hline
WD Net w/ all features & 0.769 & (0.753, 0.786) & $p<.001$ & 85.6\% & (82.6\%, 88.7\%) & $p=.004$\\
\hline
\end{tabular}
}
\end{table}

In this experiment, the best AUC value, 0.847, is achieved by merging two image-based feature, \textit{i.e.}, WLR+HLQ (line 6). With 70\% PPV, WLR+HLQ obtained a sensitivity of 92.6\%, a specificity of 37.0\% and an accuracy of 71.7\%. The best sensitivity (with PPV=70\%), 94.9\% is obtained by WLR features. Although the sensitivity of WLR is higher than WLR+HLQ, there is no significant difference ($p=0.083>0.05$). 
Table~\ref{tab:feature_rank_siteB} lists the 12 WLR+HLQ features used to obtain the best results. Table~\ref{tab:mgh_mth} shows the results of different methods on Site B dataset. Three traditional machine learning methods and the DNN (small) model used the 12 selected WLR+HLQ+DVB features (listed in Table~\ref{tab:feature_rank_siteB_all}) for prediction. SVM achieved the best AUC and sensitivity. Random forest obtained competitive AUC value with no significant difference ($p>0.05$) but inferior sensitivity ($p<0.05$). The DNN (small) model here also achieved an AUC value comparable with the best result but much lower sensitivity.

\begin{table}[h!]
\centering
\caption{The 12 best WLR+HLQ+DVB features  used for the experiments in Table~\ref{tab:mgh_mth} ranked by feature ranking strategy introduced in Section~\ref{sec:prediction} . The third and sixth columns show the Gini importance of the corresponding feature averaged in the 5-fold cross validation.}
\label{tab:feature_rank_siteB_all}
\scalebox{0.7}{
\begin{tabular}{|l|l|l||l|l|l|}
\hline
\textbf{\#} & \textbf{Quantitative features} & \textbf{G(\%)} & \textbf{\#} & \textbf{Quantitative features} & \textbf{G(\%)} \\
\hline
    1 & {\color{ForestGreen}Lobe\#2 RPO HU3\tnote{**}} & 9.90 & 2 & {\color{red}WBC}& 8.24 \\
    \hline
    3 & {\color{ForestGreen}Lobe\#2 RPO HU2} & 8.85 & 4 & {\color{ForestGreen}Whole Lung RPO HU4}& 8.78 \\
    \hline
    5  & {\color{blue} LoG($\sigma=4.5$)-Firstorder-Kurtosis \tnote{***}}& 7.6 & 6 & {\color{blue}Exponential-GLRLM-ShortRunLowGL}& 7.79 \\
    \hline
    7 & {\color{blue}Exponential-GLRLM-ShortRun} & 7.48 & 8 & {\color{blue}Sqrt-NGTDM-Strength}& 9.17 \\
    \hline
    9 & {\color{ForestGreen}LoG($\sigma=3.5$)-GLSZM-ZoneEntropy}& 7.91 & 10 & {\color{blue}Right Lung RPO HU4}& 8.0 \\
    \hline
    11 & {\color{blue}LLL-NGTDM-Strength}& 7.79 & 12 & {\color{blue}LoG($\sigma=4.5$)-NGTDM-Contrast}& 8.51 \\
    \hline
\end{tabular}
}
\begin{tablenotes}
	\footnotesize
	\item[\red{*}] \red{Red text} indicates non-imaging features.
	\item[{\color{ForestGreen}**}] {\color{ForestGreen}Green text} indicates lobe-wise quantification features, HU1-HU4 are the four HU intervals.
	\item[\blue{***}] \blue{Blue text} indicates whole lung radiomics features encoded as Filter-FeatureType-Parameter.
\end{tablenotes}
\end{table}

\subsection{Results on Site C Dataset}
\label{sec:exp_novara}

\begin{table}[h!]
\centering
\caption{Comparison among the features used in exist state-of-the-art works and different combinations of the proposed features on ICU admission prediction on Site C dataset.}
\label{tab:novara_res}
\scalebox{0.72}{
\begin{tabular}{|l|c|c|c|c|c|c|c|}
\hline
\multirow{2}{*}{\textbf{Features}} & \multicolumn{3}{c|}{\textbf{AUC}} &  \multicolumn{3}{c|}{\textbf{Sensitivity (PPV$=$70\%)}} & \multirow{2}{*}{\textbf{K}} \\
\cline{2-7}
& Mean & 95\% CI & p Value & Mean & 95\% CI & p Value &\\
\hline
Img feature (Tang 2020) & 0.763 & (0.670, 0.856) & $p=.044$ & 85.0\% & (76.4\%, 93.6\%) & $p=.020$ & 10\\
\hline
Img feature (Zhu 2020) & 0.675 & (0.645, 0.706) & $p<.001$ & 73.9\% & (59.0\%, 88.8\%) & $p<.011$ & 39\\
\hline
\hline
DVB & 0.595 & (0.524, 0.665) & $p<.001$ & 63.3\% & (36.1\%, 90.5\%) & $p=.019 $ &  4\\
\hline
HLQ & 0.691 & (0.660, 0.722) & $p<.001$ & 86.7\% & (80.3\%, 93.0\%) & $p<.014$  &  7\\
\hline
WLR & 0.815 & (0.782, 0.848) &$p=.020$ & \emph{95.6\%} & (92.8\%, 98.3\%) & $p=.186$ & 12\\
\hline
WLR+HLQ & 0.826 & (0.813, 0.839) & $p=.191$ & \textbf{96.1\%} & (94.4\%, 97.8\%) & -  & 20\\
\hline
WLR+DVB & \emph{0.835} & (0.809, 0.861) & $p=.365$ & 95.0\% & (91.6\%, 98.4\%) & $p=.088$ & 15\\
\hline
HLQ+DVB & 0.760 & (0.705, 0.815) & $p=.016$ & 85.6\%  & (77.6\%, 93.5\%) & $p<.010$ & 2\\
\hline
WLR+HLQ+DVB & \textbf{0.840} & (0.804, 0.876) & - & 94.4\% & (92.3\%, 96.6\%) & $p=.035$ & 35\\
\hline
\end{tabular}}
\end{table}

\begin{figure}[h!]
	\centering
	\includegraphics[width=1\textwidth,trim=50 20 70 50, clip]{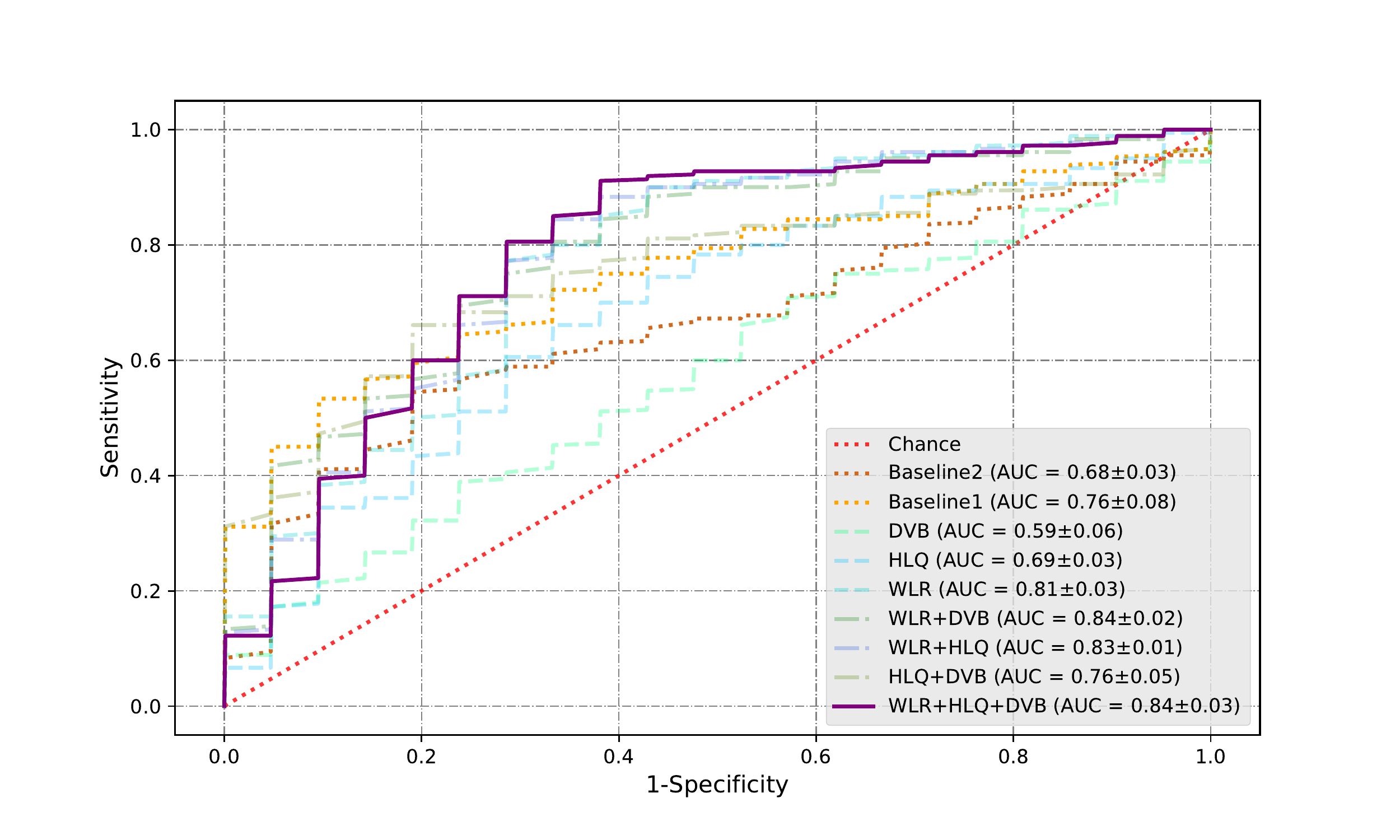}
	
	\caption{ROC curves on Site C dataset.}\label{fig:novara_auc}
\end{figure}

\begin{table}[hbt]
\centering
\caption{Comparison of different machine learning methods with selected features on Site C dataset.}
\label{tab:inovara_mths}
\scalebox{0.72}{
\begin{tabular}{|l|c|c|c|c|c|c|}
\hline
\multirow{2}{*}{\textbf{Methods}} & \multicolumn{3}{c|}{\textbf{AUC}} &  \multicolumn{3}{c|}{\textbf{Sensitivity (PPV$=$70\%)}} \\
\cline{2-7}
& Mean & 95\% CI & p Value & Mean & 95\% CI & p Value\\
\hline
Random Forests & \textbf{0.840} & (0.804, 0.876) & - &  \textbf{94.4\%} & (92.3\%, 96.6\%) & -  \\
\hline
SVM & 0.811 & (0.782, 0.839) & $p=.031$ & 93.3\% & (89.2\% , 97.5\%) & $p=.324$ \\
\hline
Logistic Regression & 0.717 & (0.666, 0.768) & $p=.009$ & 86.7\% & (79.3\%, 94.0\%) & $p=.019$\\
\hline
\hline
DNN (small) & 0.695 & (0.619, 0.771) & $p=.011$ & 77.8\% & (53.6\%, 100.0\%) & $p=.082$\\
\hline
DNN w/ all features & 0.568 & (0.550, 0.587) & $p<.001$ & 52.8\% & (29.2\%, 76.4\%) & $p=.006$ \\
\hline
WD Net w/ all features & 0.528 & (0.479, 0.578) & $p<.001$ & 34.4\% & (24.1\%, 44.8\%) & $p<.001$\\
\hline
\end{tabular}
}
\end{table}

Results on Site C dataset are shown in Table~\ref{tab:novara_res} and Fig.~\ref{fig:mgh_auc}. The non-imaging (DVB) features alone also didn't achieve well performance. It might be because many petients' DVB features are missing or incomplete as shown in Table.~\ref{tab:feature_novara} Yet, the introduce of DVB features significantly improves the AUC performance of HLQ features ($p=0.014<0.05$). In this experiment, the best AUC value, 0.840, is achieved by merging all three kinds of features. While maintaining a PPV of 70\%, it achieved a sensitivity of 94.4\%, a specificity of 33.3\% and an accuracy of 71.9\%. Table~\ref{tab:feature_rank_siteC} shows the 35 features used to obtain the best results at the end of this paper due to its large size. A comparison of different methods on Site C dataset is shown in Table~\ref{tab:inovara_mths}. Random forest obtained the best AUC and sensitivity. The performance of all three deep learning networks is less than adequate. One of the most important reasons may be that Site C dataset is considerably smaller than the other two datasets (only containing 57 cases). Compared with traditional machine learning methods, the number of parameters in deep learning models is several orders of magnitude larger, which makes them much more vulnerable to overfitting with a limited training set.

\subsection{Generalization Ability}
In this section, we further evaluate if feature combinations learned from one site can be generated to other sites. Experiments on all 6 permutations are conducted (train on Site A, test on Site B; train on Site A, test on Site C; train on Site B, test on Site A; train on Site B, test on Site C; train on Site C, test on Site A; train on Site C, test on Site B). Considering the 3 datasets contains differet DVB features, WLR+HLQ features are used in this section. The results are shown in Table~\ref{tab:gen_res}.

\begin{table}[h!]
\centering
\caption{Transferring WLR+HLQ features across the three datasets.}
\label{tab:gen_res}
\scalebox{0.7}{
\begin{tabular}{|c|c|c|c|c|c|c|c|}
\hline
\textbf{Methods} & \textbf{A $\to$ B} & \textbf{A $\to$ C}
 & \textbf{B $\to$ A} & \textbf{B $\to$ C}
 & \textbf{C $\to$ A} & \textbf{C $\to$ B} & \textbf{Mean}\\
\hline
Random Forests & 0.740 (36) & 0.685 (29) 
 & 0.754 (12) & 0.633 (11) 
 & 0.591 (17) & 0.717 (20) & 0.687\\
\hline
SVM & \textbf{0.774 (1)} & \textbf{0.686 (23)}
 & \textbf{0.777 (3)} & \textbf{0.710 (1)}
 & \textbf{0.649 (17)} & 0.694 (3) & 0.715\\
\hline
Logistic Regression & 0.744 (11) & 0.706 (23)
 & 0.756 (1) & 0.698 (1)
 & 0.642 (19) & \textbf{0.752 (3)} & 0.716\\
\hline
\end{tabular}}
\end{table}

There were tremendous differences in the geographic distribution and scanner technologies used for imaging patients at the three participating sites. Despite this, we achieved AUC values as high as 0.777 for WLR+HLQ features. Some variations in the AUCs and performance of our model across different sites is expected due to challenges associated with acquisition of consistent data variables and practices. 
The results in Table~\ref{tab:gen_res} shows that SVM and logistic regression achieved very similar performance on average with no significant difference ($p=0.462$).Although random forest did not perform well in this transfer experiment, the difference between random forest and SVM was not significant ($p=0.058>0.05$) either. Considering that Random Forest outperformed logistic regression on all three datasets as presented in Section~\ref{sec:exp_iran}-\ref{sec:exp_novara}, it is still the overall best performing method in our study.

Our study stresses the need to combine imaging findings with clinical, laboratory and management variables which can improve the model performance, aid in better performance statistics on each dataset. On the other hand, complexities of disease and its outcomes are tied to local factors and stress the importance of tweaking the best models based on rich local or institutional level factors rather than a single “one-type-fit-all” model.

\section{Discussion and Conclusions}

In this paper, we propose to combine size and volume information of the lungs and manifestations, radiomics features of pulmonary opacities and non-imaging DVB features to predict need for ICU admission in patients with COVID-19 pneumonia. Metrics related to ICU admission rates, need and availability are key markers in management of individual patients as well as in resource planning for managing high prevalence diseases. To the best of our knowledge, this is the first study that uses holistic information of a patient including both imaging and non-imaging data to predict patient outcome.

Although promising results were achieved, the study has a few limitations. First of all, due to the limited size of our datasets, we could not conduct more fine-grained outcome predictions. 
The size of the available datasets could also be the reason that more complex models, such as deep neural networks, did not perform well in our experiments.
Once larger datasets are available, our model can be rapidly adapted to assess generalization ability and to establish implications on datasets from other sites. Efforts are underway (such as within Radiological Society of North America) to establish such imaging datasets of COVID-19 pneumonia. 
Second, variations in performance of different imaging and clinical features on datasets from three sites underscore the need for careful local vetting of deep learning predictive models. Future models should take into account regional bias introduced from different criteria on imaging use, underlying patient comorbidities, and management strategies, so that more robust models can be built. This also goes beyond the generalization ability of machine learning in medical applications. The best and most relevant results likely require regional, local or even site-specific tuning of predictive models. This is especially true in context of the three sites, which are under very different healthcare systems as in our study. We believe that this limitation is not unique to our model. Last but not the least, another limitation of our study pertains to the lack of  access to the specific treatment regimens at the three sites; their inclusion could have further enhanced the accuracy of our algorithm.
However, it also suggests that this generic approach can be trained on data from a hospital to create a customized predictive model for clinical decision support.

In summary, our integrative analysis machine learning based predictive model can help assess disease burden and forecast meaningful patient outcomes with high predictive accuracy in patients with COVID-19 pneumonia. Many patients with adverse outcomes from COVID-19 pneumonia and cardiorespiratory failure develop diffuse alveolar damage and adult respiratory distress syndrome (ARDS), which are also well-known end stage manifestations of other pulmonary diseases such as from other infections and lung injuries. Although we did not test our model in patients with ARDS from non-COVID causes, given the overlap in imaging and clinical features of respiratory failure, we expect that the methods of quantifying pulmonary opacities used in our approach will extend beyond COVID-19 pneumonia. 
In addition, introducing data of diseases with similar properties with COVID-19 may further improve the robustness and performance of our approach, which will be explored in our future work. Further studies will help assess such applications beyond the current pandemic of COVID-19 pneumonia.

\begin{table}[t]
	\centering
	\caption{The top 52 features ranked by feature ranking strategy introduced in Section~\ref{sec:prediction} on Site A dataset. The third and sixth columns show the Gini importance of the corresponding feature averaged in the 5-fold cross validation.}
	\label{tab:feature_rank_siteA}
	\scalebox{0.8}{
	\begin{tabular}{|l|l|l||l|l|l|}
	\hline
	\textbf{\#} & \textbf{Quantitative features} & \textbf{G(\%)} & \textbf{\#} & \textbf{Quantitative features} & \textbf{G(\%)} \\
	\hline
	1 & {\color{red} L/W ratio\tnote{*}} & 7.83 & 2 & {\color{blue}Sqrt-NGTDM-Strength\tnote{**}}& 2.68 \\
	\hline
	3 & {\color{blue}LLL-GLDM-LowGray} & 2.15 & 4 & {\color{ForestGreen}Lobe\#5 RPO HU4\tnote{***}}& 2.27 \\
	\hline
	5  & {\color{blue}Org-GLDM-LowGray}& 1.85 & 6 & {\color{ForestGreen}Lobe\#5 RPO HU3}& 2.03 \\
	\hline
	7 & {\color{blue}LLL-GLRLM-LongRunLG} & 1.71 & 8 & {\color{ForestGreen}Left Lung RPO HU4}& 2.16 \\
	\hline
	9 & {\color{red}Lym count}& 2.62 & 10 & {\color{ForestGreen}Lobe\#2 VPO HU3}& 1.7 \\
	\hline
	11 & {\color{ForestGreen}Lobe\#2 VPO HU2}& 1.78 & 12 & {\color{blue}LLL-GLRLM-LGRun}& 2.23 \\
	\hline
	13 & {\color{blue}LoG($\sigma=1.5$)-GLCM-MCC} & 3.04 & 14 & {\color{blue}LLL-GLSZM-LGZone}& 1.8 \\
	\hline
	15 & {\color{blue}Org-NGTDM-Strength} & 1.71 & 16 & {\color{ForestGreen}Lobe\#5 VPO HU3}& 1.82 \\
	\hline
	17 & {\color{blue}Org-GLRLM-ShortRunLG} & 1.34 & 18 & {\color{blue}LLL-NGTDM-Strength}& 1.63 \\
	\hline
	19 & {\color{blue}Org-GLRLM-LowGrayRun} & 1.43 & 20 & {\color{blue}LLL-GLDM-LargeDepdLG}& 1.94 \\
	\hline
	21 & {\color{blue}LLL-GLRLM-ShortRunLG} & 1.73 & 22 & {\color{blue}LLL-GLSZM-SmallAreaLG}& 1.39 \\
	\hline
	23 & {\color{blue}LLL-GLSZM-SmallAreaHG} & 1.93 & 24 & {\color{blue}LoG($\sigma=2.5$)-GLCM-MCC}& 2.62 \\
	\hline
	25 & {\color{red}WBC} & 1.81 & 26 & {\color{blue}Org-Shape-LeastAxisLength}& 1.57 \\
	\hline
	27 & {\color{blue}Org-GLRLM-LongRunLG} & 1.22 & 28 & {\color{blue}LoG($\sigma=1.5$)-GLCM-Corr}& 2.54 \\
	\hline
	29 & {\color{ForestGreen}Lobe\#2 Infection Ratio HU4} & 1.4 & 30 & {\color{blue}HLL-FirstOrder-Mean}& 1.91 \\
	\hline
	31 & {\color{ForestGreen}Left Lung VPO HU2} & 1.51 & 32 & {\color{blue}LoG($\sigma$=2.5)-GLSZM-LALG}& 2.12 \\
	\hline
	33 & {\color{blue}LLL-GLCM-JointAverage} & 1.45 & 34 & {\color{blue}Logarithm-NGTDM-Strength}& 1.43 \\
	\hline
	35 & {\color{ForestGreen}Lobe\#2 VPO HU4} & 1.18 & 36 & {\color{blue}HLH-GLSZM-Zone\%}& 2.83 \\
	\hline
	37 & {\color{blue}LoG($\sigma=2.5$)-GLCM-Corr} & 1.61 & 38 & {\color{blue}LLL-GLSZM-HGZone}& 1.29 \\
	\hline
	39 & {\color{ForestGreen}Lobe\#4 VPO HU4} & 2.02 & 40 & {\color{blue}LLL-GLCM-SumAverage}& 1.19 \\
	\hline
	41 & {\color{blue}LoG($\sigma=2.5$)-NGTDM-Busyness} & 1.47 & 42 & {\color{blue}LLH-FirstOrder-Skewness}& 1.83 \\
	\hline
	43 & {\color{ForestGreen}Lobe\#2 RPO HU3 } & 1.32 & 44 & {\color{blue}LoG($\sigma=1.5$)-GLSZM-ZonVar}& 2.07 \\
	\hline
	45 & {\color{blue}Sqrt-GLRLM-RunLenNonUniform} & 1.29 & 46 & {\color{ForestGreen}Lobe\#5 VPO HU4}& 0.88 \\
	\hline
	47 & {\color{blue}LoG($\sigma=2.5$)-GLCM-Imc2} & 1.86 & 48 & {\color{blue}LoG($\sigma=1.5$)-GLSZM-LALG}& 2.19 \\
	\hline
	49 & {\color{blue}HHH-GLCM-SumAverage} & 1.89 & 50 & {\color{ForestGreen}Left Lung VPO HU3}& 1.39 \\
	\hline
	51 & {\color{blue}HHH-GLCM-JointAverage} & 2.07 & 52 & {\color{ForestGreen}Left Lung RPO HU3}& 1.29 \\
	\hline
	\end{tabular}
	}
	\begin{tablenotes}
		\footnotesize
		\item[\red{*}] \red{Red text} indicates non-imaging features.
		\item[{\color{ForestGreen}**}] {\color{ForestGreen}Green text} indicates lobe-wise quantification features, HU1-HU4 are the four HU intervals.
		\item[\blue{***}] \blue{Blue text} indicates whole lung radiomics features encoded as Filter-FeatureType-Parameter.
	\end{tablenotes}
	
\end{table}

\begin{table}[hbt]
	\centering
	\caption{The top 35 features ranked by feature ranking strategy introduced in Section~\ref{sec:prediction} on Site C dataset. The third and sixth columns show the Gini importance of the corresponding feature averaged in the 5-fold cross validation.}
	\label{tab:feature_rank_siteC}
	\scalebox{0.7}{
	\begin{tabular}{|l|l|l||l|l|l|}
	\hline
	\textbf{\#} & \textbf{Quantitative features} & \textbf{G(\%)} & \textbf{\#} & \textbf{Quantitative features} & \textbf{G(\%)} \\
	\hline
	1 & {\color{blue} LoG($\sigma=2.5$)-GLRLM-LongRunHighGL \tnote{*}} & 4.39 & 2 & {\color{ForestGreen}Lobe\#2 VPO HU3\tnote{**}}& 3.2 \\
    \hline
    3 & {\color{blue} Original-Shape-LeastAxisLength} & 2.99 & 4 & {\color{blue} LoG($\sigma=2.5$)-NGTDM-Complexity}& 2.92 \\
    \hline
    5 & {\color{ForestGreen}Lobe\#2 RPO HU4} & 3.24 & 6 & {\color{ForestGreen}Lobe\#2 RPO HU3}& 3.4 \\
    \hline
    7 & {\color{ForestGreen}Lobe\#3 RPO HU1} & 2.91 & 8 & {\color{blue} LoG($\sigma=2.5$)-GLRLM-ShortRunLowGL}& 2.88 \\
    \hline
    9 & {\color{blue} LLH-Firstorder-Mean} & 3.99 & 10 & {\color{ForestGreen}Lobe\#2 VPO HU4}& 2.61 \\
    \hline
    11 & {\color{blue} LoG($\sigma=3.5$)-firstorder-Range} & 2.19 & 12 & {\color{blue} LoG($\sigma=3.5$)-GLDM-LowGL}  & 2.17 \\
    \hline
    13 & {\color{blue} LoG($\sigma=3.5$)-GLCM-Corr} & 3.91 & 14 & {\color{red} Age\tnote{***}}& 2.72 \\
    \hline
    15 & {\color{blue} LoG($\sigma=2.5$)-GLRLM-LowGLRun} & 1.76 & 16 & {\color{blue} LoG($\sigma=4.5$)-Firstorder-RMS}& 2.96 \\
    \hline
    17 & {\color{blue} LoG($\sigma=4.5$)-GLCM-MaxProb} & 3.34 & 18 & {\color{blue} LoG($\sigma=4.5$)-GLCM-Idmn}& 2.36 \\
    \hline
    19 & {\color{blue} HLL-Firstorder-Mean} & 3.1 & 20 & {\color{blue} LoG($\sigma=2.5$)-GLDM-SmallDepLowLG}& 2.94 \\
    \hline
    21 & {\color{blue}LoG($\sigma=4.5$)-FirstOrder-Range} & 2.05 & 22 & {\color{blue}LoG($\sigma=2.5$)-GLDM-LargeDepLowLG} & 3.31 \\
    \hline
    23 & {\color{blue}LHL-FirstOrder-Maximum} & 2.23 & 24 & {\color{blue}LoG($\sigma=2.5$)-GLCM-LDMN} & 3.18 \\
    \hline
    25 & {\color{blue}LoG($\sigma=1.5$)-NGTDM-Strength} & 3.05 & 26 & {\color{blue}LoG($\sigma=4.5$)-NGTDM-Complexity} & 2.09 \\
    \hline
    27 & {\color{blue}LoG($\sigma=2.5$)-GLSZM-LowGray} & 1.79 & 28 & {\color{blue}LHL-FirstOrder-Kurtosis} & 2.98 \\
    \hline
    29 & {\color{blue}LoG($\sigma=1.5$)-GLCM-MCC} & 3.29 & 30 & {\color{blue}LoG($\sigma=3.5$)-GLCM-IMC2} & 2.88 \\
    \hline
    31 & {\color{blue}Logarithm-FirstOrder-Energy} & 3.38 & 32 & {\color{blue} Exponential-FirstOrder-Energy} & 2.6 \\
    \hline
    33 & {\color{blue}LoG($\sigma=2.5$)-GLCM-SumAverage} & 2.03 & 34 & {\color{blue}LoG($\sigma=2.5$)-GLCM-LDN} & 2.3 \\
    \hline
    35 & {\color{blue}LLH-FirstOrder-Kurtosis} & 2.87 & &&\\
    \hline
	\end{tabular}
	}
	\begin{tablenotes}
		\footnotesize
		\item[\red{*}] \red{Red text} indicates non-imaging features.
		\item[{\color{ForestGreen}**}] {\color{ForestGreen}Green text} indicates lobe-wise quantification features, HU1-HU4 are the four HU intervals.
		\item[\blue{***}] \blue{Blue text} indicates whole lung radiomics features encoded as Filter-FeatureType-Parameter.
	\end{tablenotes}
	
\end{table}

\section*{Acknowledgments}
This work was partially supported by National Institute of Biomedical Imaging and Bioengineering (NIBIB) under award R21EB028001 and National Heart, Lung, and Blood Institute (NHLBI) under award R56HL145172.

\newpage

\bibliographystyle{model2-names}
\biboptions{authoryear}
\bibliography{refs}

\end{document}